\documentclass[twocolumn,aps,showpacs]{revtex4}
\usepackage{graphicx}
\def\PRL{Phys. Rev. Lett. }
\def\PRC{Phys. Rev. C}
\def\PRD{Phys. Rev. D}
\def\PLB{Phys. Lett. B}

\def\etal{\emph{et al.}}
\begin{document}

\title{Disappearance of back-to-back high $p_T$ hadron correlations in 
central Au+Au collisions at $\sqrt{s_{NN}}$ = 200 GeV}

\author{
C.~Adler$^{11}$, Z.~Ahammed$^{23}$, C.~Allgower$^{12}$, J.~Amonett$^{14}$,
B.D.~Anderson$^{14}$, M.~Anderson$^5$, G.S.~Averichev$^{9}$, 
J.~Balewski$^{12}$, O.~Barannikova$^{9,23}$, L.S.~Barnby$^{14}$
% ...
\\(STAR Collaboration)
\\
% ...
$^5$University of California, Davis, California 95616
\\
% ...
$^{9}$Laboratory for High Energy (JINR), Dubna, Russia
\\
% ...
$^{11}$University of Frankfurt, Frankfurt, Germany
\\$^{12}$Indiana University, Bloomington, Indiana 47408
\\
% ...
$^{14}$Kent State University, Kent, Ohio 44242
\\
% ...
$^{23}$Purdue University, West Lafayette, Indiana 47907
\\
% ...
}

\begin{abstract}

Azimuthal correlations for large transverse momentum 
charged hadrons have been
measured over a wide pseudo-rapidity range and full azimuth in Au+Au
and p+p collisions at $\sqrt{s_{NN}}$ = 200 GeV.    The
small-angle correlations observed in p+p collisions 
and at all centralities of Au+Au collisions 
are characteristic of hard-scattering processes already observed in 
elementary collisions.  A strong back-to-back
correlation exists for p+p and peripheral Au + Au. In 
contrast, the back-to-back correlations are
reduced considerably in the most central Au+Au collisions, indicating
substantial interaction as the hard-scattered partons or their fragmentation
products traverse the medium.

\end{abstract}
\maketitle

In collisions of heavy nuclei at high energies, a new state of matter
consisting of deconfined quarks and gluons at high density is expected
\cite{QGP}.  Large transverse momentum partons in the high-density
system result from the initial hard scattering of nucleon constituents.
After a hard scattering, the parton fragments to create a high energy
cluster (jet) of particles.  A high momentum parton traversing a dense
colored medium is predicted to experience substantial energy loss
\cite{Gyulassy, Wang} and may be absorbed.  Measurement of the parton
fragmentation products after hard-scattering processes in nuclear
collisions may reveal effects due to the interaction of
high-momentum partons traversing the medium, thereby measuring the gluon 
density of the medium \cite{Baier}.

Hard scattering processes have been established at high transverse
momentum ($p_T$) in elementary collisions at high energy through the
measurement of jets \cite{UA2_jets, UA1_jets, CDF_jets}, correlated
back-to-back jets (di-jets) \cite{CDF_dijets}, high $p_T$ single
particles, and back-to-back correlations between high $p_T$ hadrons
\cite{highpt_corr}.  Jets have been shown to carry the momentum of the
parent parton \cite{fragmentation}.  The jet cross sections and high
$p_T$ single particle spectra are well described over a broad range of
energies \cite{QCD} in terms of the hadron's parton distributions, hard
parton scattering treated by perturbative QCD, and subsequent
fragmentation of the parton.  High $p_T$ jet events have also been
studied in proton-nucleus interactions \cite{pA_jets}.  In the absence
of effects of the nuclear medium the rate of hard processes should scale
linearly with the number of binary nucleon-nucleon collisions.  Recent
results from RHIC, however, show a suppression of the single
particle inclusive spectra of hadrons for $p_T > 2$ GeV/c in central
Au+Au collisions, indicating substantial in-medium interactions
\cite{PHENIX_highpt, STAR_highpt}.

In this Letter, we report measurements of two-hadron angular
correlations at large transverse momentum for p+p and Au+Au collisions
at $\sqrt{s_{NN}} = 200 $ GeV.  These correlation measurements provide
the most direct evidence for production of jets in high energy
nucleus-nucleus collisions, and allow for the first time measurements,
inaccessible in inclusive spectra, of the
fate of back-to-back jets in the dense medium as a function of the size
of the overlapping system.  The results reveal significant interaction of hard-scattered partons (or their fragmentation products) in the medium, with a strong dependence on the geometry and distance of traversal. 

The measurements were made using the STAR detector \cite {STARNIM} at
the Relativistic Heavy-Ion Collider (RHIC) at Brookhaven National
Laboratory.  The STAR detector is a large acceptance magnetic
spectrometer, with a large volume Time Projection Chamber (TPC) inside a
0.5 Tesla solenoidal magnet.  The TPC measures the trajectories of
charged particles and determines the particle momenta.  The TPC has full
azimuthal coverage over a pseudo-rapidity range $|\eta|<1.5$.  STAR has
excellent position and momentum resolution, and, due to its vertexing
capabilities, is able to identify many sources of secondary particles.
The p+p analysis uses $\approx$10 million minimum bias p+p events
triggered on the coincidence of signals from scintillator annuli
spanning the pseudo-rapidity interval $3.5\le |\eta| \le 5.0$.  The
Au+Au analysis uses $\approx$1.7 million minimum bias Au+Au events and
$\approx$1.5 million top 10\% central Au+Au events.

Partons fragment into jets of hadrons in a cone around the direction of
the original hard-scattered parton.  The leading hadron in the jet tends
to be most closely aligned with the original parton direction
\cite{UA1_83}.  The large multiplicities in Au+Au collisions make full
jet reconstruction impractical.  Thus, we utilize two-particle azimuthal
correlations of high $p_T$ charged hadrons \cite{STARv2} to identify
jets on a statistical basis, with known sources of background
correlations subtracted.

Events with at least one large transverse momentum hadron
($4<p_T^{trig}<6$ or $3<p_T^{trig}<4$ GeV/c), defined to be a {\it trigger}
particle, are used in this analysis.  
For each of the trigger particles in the event, we increment the number
$N(\Delta \phi, \Delta \eta)$ of {\it associated} tracks with 2 GeV/c $<p_T<p_T^{trig}$ as a function of their azimuthal ($\Delta \phi$) and pseudo-rapidity
($\Delta \eta$) separations from the trigger particle.  We then construct
an overall azimuthal pair distribution per trigger particle, 
\begin{equation} 
D(\Delta \phi) \equiv \frac{1}{N_{trigger}}\frac{1}{\epsilon} \int d\Delta
\eta N(\Delta \phi, \Delta \eta),
\end{equation} 
where $N_{trigger}$ is the
observed number of tracks satisfying the trigger requirement. 
The efficiency $\epsilon$ 
for finding the associated particle is evaluated by embedding simulated
tracks in real data.  In order to have a high and constant tracking
efficiency, the tracks are required to have $|\eta|<0.7$, which
translates to a relative pseudo-rapidity acceptance of $|\Delta
\eta|<1.4$.  The single track reconstruction efficiency varies from 77\%
for the most central Au+Au collisions to 90\% for the most peripheral
Au+Au collisions and p+p collisions.

\begin{figure}[tb]
\begin{center}
\includegraphics[height=5cm]{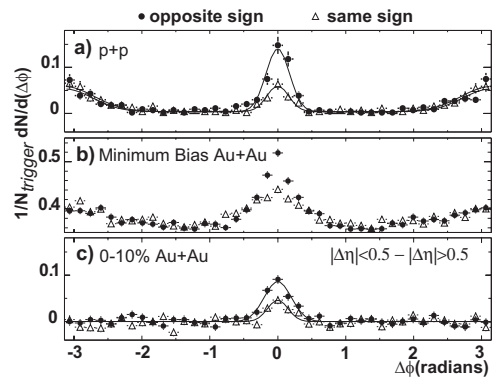}
\end{center}
\caption{Azimuthal distributions of same-sign
and opposite-sign pairs for a) p+p, b) minimum bias Au+Au,
and c) background-subtracted
central Au+Au collisions.  All
correlation functions
require a trigger particle with $4<p_T^{trig}<6$ GeV/c and
associated particles with $2$ GeV/c $<p_T<p_T^{trig}$.  The curves
are one- or two- Gaussian fits.}
\label{chargesign}
\end{figure}

Identical analysis procedures are applied to the p+p and Au+Au data.
Displayed in Figure \ref{chargesign} are the azimuthal distributions for
same-sign and opposite-sign charged pairs from the a) p+p data and b)
minimum bias Au+Au data for $4<p_T^{trig}<6$ GeV/c.  The data are
integrated over the relative pseudo-rapidity range $0 < |\Delta \eta| <
1.4$.  Clear correlation peaks are observed near $\Delta \phi \sim 0$
and $\Delta \phi \sim \pi$ in the data.  The opposite-sign correlations
at small relative azimuth are larger than those of the same-sign
particle pairs, while the sign has a negligible effect on the
back-to-back correlations.


To isolate the jet-like correlations (localized in $\Delta \phi$,
$\Delta \eta$) in central Au+Au collisions, the azimuthal distributions
are measured for two regions of relative pseudo-rapidity, $|\Delta \eta|
< 0.5 $ and $0.5 < |\Delta \eta| < 1.4$ \cite{STARv2}.  The difference
between the small and large relative pseudo-rapidity azimuthal
distributions is displayed in Figure \ref{chargesign}c
along with single Gaussian fits.  Near $\Delta \phi$ = 0, the azimuthal
distributions from Au+Au and p+p have similar shapes.  For the
opposite-sign azimuthal distributions, the Gaussian widths are
$0.17\pm0.01(stat.)\pm0.03(sys.)$ radians for p+p data, and
$0.20\pm0.02(stat.)\pm0.03(sys.)$ radians for the central Au+Au data.
For the same-sign azimuthal distributions, the Gaussian widths are
$0.16\pm0.02(stat.)\pm0.03(sys.)$ radians for p+p data, and
$0.15\pm0.03(stat.)\pm0.04(sys.)$ radians for the central Au+Au data.
The systematic errors reflect the spread
of values found for different choices of the $\Delta \phi$ bin width.
Within the errors,
there are no significant differences between the small-angle correlation widths
for p+p and central Au+Au collisions.  

The ratios of the opposite-sign to same-sign peak areas are
$2.7\pm0.9(stat.)\pm0.2(sys.)$ for p+p and
$2.5\pm0.6(stat.)\pm0.2(sys.)$ for central Au+Au collisions.  In jet
fragmentation, there are dynamical charge correlations between the
leading and next-to-leading charged hadrons
\cite{Delphi_charge_ordering} that originate from the formation of
$q\bar{q}$ pairs along a string between two partons.  This results in a
preferential ordering into oppositely-charged adjacent particles along a
string during fragmentation.  The Hijing event generator, which utilizes
the Lund string fragmentation scheme \cite{lund} incorporating these
concepts, predicts a ratio of $2.6\pm0.7$ for the opposite-sign to
same-sign correlation strengths.  The agreement of this ratio with 
those measured in the central Au+Au and p+p suggests that the same jet
production mechanism is responsible for a majority of the charged
hadrons with $p_T>4$ GeV/c in p+p and central Au+Au collisions.

The decay of resonances would also lead to small-angle azimuthal correlations, 
but a resonance decay origin is unlikely due
to the observed correlation of particles with the same charge sign, the
similarity in the measured small-angle 
azimuthal correlation widths in the Au+Au and p+p
interactions, and the strong back-to-back correlations of large $p_T$ particles
seen for p+p collisions in Fig. \ref{chargesign}a.  The latter correlations,
indicative of di-jet events \cite{highpt_corr}, are removed from the central 
Au+Au sample by the subtraction in Fig. \ref{chargesign}c. A quantitative
analysis of back-to-back jet survival in Au+Au requires the more detailed
treatment of background correlations described below.  

In addition to correlations due to jets, the two-particle
azimuthal distributions in Au+Au exhibit a structure attributable to an
anisotropy of single particle production relative to the reaction plane.
Previous measurements \cite{STARv2} indicate that, even at large
transverse momentum, the particle distributions contain an anisotropy
due to elliptic flow that can be characterized by $dN/d(\phi-\Phi_r)
\propto 1+2v_2 \cos(2(\phi-\Phi_r))$, where $\Phi_r$ is the reaction
plane angle determined event by event and $v_2$ is the elliptic flow
parameter.  This leads to a two particle azimuthal distribution of the
form, $dN/d\Delta \phi = B(1+2v_2^2 \cos(2\Delta\phi))$.  The elliptic
flow component of the two-particle azimuthal distribution is measured
using several methods \cite{STARv2}.  In this paper, $v_2$ is determined
using a reaction-plane method.  

A simple reference model can be constructed for
the two-particle azimuthal distributions of high $p_T$
particles in Au+Au collisions. 
A number of independent hard scatterings (each similar to one measured 
in a triggered p+p event) 
included in an event with correlations
due to elliptic flow can be represented by
the azimuthal distribution,
\begin{equation}
D^{\mathrm{model}} = D^{\mathrm{pp}} + B(1+2v_2^2 \cos(2\Delta \phi)).
\label{c2eqn}
\end{equation}
The elliptic flow parameter ($v_2$)  is measured independently in
the same set of events, and is taken to be constant for $p_T>2$ GeV/c
\cite{STARv2}. 
The parameter $B$ is then determined by fitting the observed 
$D^{\mathrm{AuAu}}$ in the region 
$0.75<|\Delta \phi|<2.24$ radians,
which is largely
free of jet contributions in the p+p data.   

\begin{figure}[tb]
\begin{center}
\includegraphics[width=6cm]{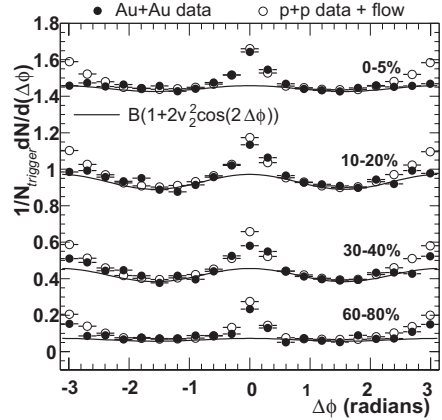}
\end{center}
\caption{Azimuthal distributions 
($0<|\Delta \eta|<1.4$, $4<p_T^{trig}<6$ GeV/c) for Au+Au 
collisions (solid circles) compared to
the expected distributions ${D^{\mathrm{model}}}$ from 
Equation \protect{\ref{c2eqn}} (open circles).  
Also shown is the elliptic flow contribution for each
centrality (solid curve).}
\label{data}
\end{figure}

In Figure \ref{data}, the azimuthal distributions for $0<|\Delta \eta|<1.4$ 
in Au+Au
collisions at various centralities are compared to Equation \ref{c2eqn}
using the measured p+p data.  The centrality selection is constructed
by subdividing the Au+Au minimum bias
data sample into subsamples with different
charged particle multiplicities
within $|\eta| < 0.5$.
The parameters $v_2$ and $B$ are determined independently for each centrality
bin, and are listed in Table \ref{thetable}.  For all
centralities, the azimuthal correlation near $\Delta \phi = 0$ is well
described by Equation \ref{c2eqn}.  
This indicates that
the same mechanism (hard parton scattering and fragmentation)
is responsible for high transverse momentum particle production in p+p
and Au+Au collisions.  However, the
back-to-back correlations are suppressed in Au+Au collisions compared to the
expectation from Equation \ref{c2eqn}, and
the suppression is greater for more central collisions.  The most
central collisions show no indication of any back-to-back
correlations beyond that expected from elliptic flow.

\begin{table}[tb]
\centering
\begin{tabular}{|c|c|c|c|} \hline
Centrality (\%) & $N_{part}$ & $v_2$  & $B$ \\ \hline
60-80 & 20$\pm$6    & 0.24$\pm$0.04 & 0.065$\pm$ 0.003 \\ 
40-60 & 61$\pm$10   & 0.22$\pm$0.01 & 0.231$\pm$ 0.003 \\ 
30-40 & 114$\pm$13  & 0.21$\pm$0.01 & 0.420$\pm$ 0.005 \\ 
20-30 & 165$\pm$13  & 0.19$\pm$0.01 & 0.633$\pm$ 0.005 \\ 
10-20 & 232$\pm$11  & 0.15$\pm$0.01 & 0.931$\pm$ 0.006 \\ 
5-10  & 298$\pm$10  & 0.10$\pm$0.01 & 1.187$\pm$ 0.008 \\ 
0-5   & 352$\pm$7   & 0.07$\pm$0.01 & 1.442$\pm$ 0.003 \\ \hline
\end{tabular}
\caption{Centrality, number of participants, $v_2$ ($2<p_T<6$ GeV/c), and
normalization constant $B$.  The errors on $v_2$ and $B$ are statistical 
only, while the errors on the number of participants are systematic 
\protect{\cite{STAR_highpt}}.}
\label{thetable}
\end{table} 

\begin{figure}[tb]
\begin{center}
\includegraphics[height=5cm]{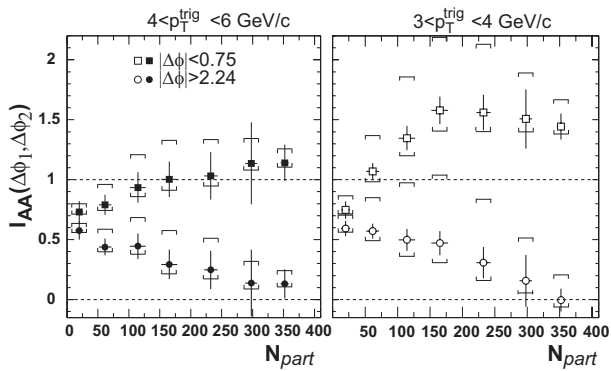}
\end{center}
\caption{Ratio of Au+Au and p+p (Equation \protect{\ref{ratioeqn}}) 
for small-angle (squares, $|\Delta \phi|<0.75$ radians) and 
back-to-back (circles, $|\Delta \phi|>2.24$ radians) azimuthal regions 
versus number of participating nucleons 
for trigger particle intervals $4<p_T^{trig}<6$ GeV/c (solid) and
$3<p_T^{trig}<4$ GeV/c (hollow). The horizontal bars indicate the 
dominant systematic error (highly correlated among points) due
to the uncertainty in $v_2$.}
\label{suppress}
\end{figure}

The ratio of the measured Au+Au correlation excess 
relative to the p+p correlation is:
\begin{eqnarray}
\lefteqn{I_{AA}(\Delta \phi_1,\Delta \phi_2) = } \nonumber\\
& \frac{\int_{\Delta \phi_1}^{\Delta \phi_2} d(\Delta \phi) [D^{\mathrm{AuAu}}- B(1+2v_2^2 \cos(2 \Delta \phi))]}{\int_{\Delta \phi_1}^{\Delta \phi_2} d(\Delta \phi) D^{\mathrm{pp}}}.
\label{ratioeqn}
\end{eqnarray}
The ratio can be plotted as a function of
the number of participating nucleons ($N_{part}$), deduced from the centrality
bins as described
in reference \cite{STAR_highpt}.  $I_{AA}$ is measured for both the small-angle ($|\Delta \phi|<0.75$ radians) and back-to-back ($|\Delta \phi|>2.24$ radians) regions.
The ratio should be unity
if the hard-scattering component of
Au+Au collisions is simply a superposition of p+p collisions unaffected
by the nuclear medium.  
These ratios are given in Figure \ref{suppress} for the trigger 
particle momentum ranges indicated.
The asymmetric systematic errors are dominated by the +5/-20\% systematic
uncertainty on $v_2$ due to the potential non-flow
contributions \cite{STARcumulant} as well as other sources of systematic uncertainty \cite{STARv2}.   

For the most peripheral bin (smallest $N_{part}$), both the 
small-angle and back-to-back
correlation strengths are suppressed compared to the expectation
from Equation \ref{c2eqn}.  This may be an indication
of initial state nuclear effects such as shadowing of parton distributions
or scattering by multiple nucleons, or
may be indicative of energy loss in a dilute medium \cite{WWA}.
As $N_{part}$ increases, 
the small-angle correlation strength increases, with 
a more pronounced increase 
for the trigger particles with lower $p_T$ threshold.  
If there were a large non-jet
contribution to particle production (i.e. collective transverse flow) 
at the trigger threshold and above, it would dilute the jet related correlation
signal and this ratio 
would be reduced.  
The back-to-back correlation strength, above background from elliptic flow,
decreases with increasing $N_{part}$ and 
is consistent with zero for the most central collisions.
In the extreme case, if there were \emph{no} elliptic flow for the 
0-5\% most central 
collisions, $I_{AA}(2.24,\pi)=0.4\pm0.1$ for $4<p_T^{trig}<6$ GeV/c, 
compared to $I_{AA}(2.24,\pi)=0.1\pm0.1$ using the 
measured elliptic flow value. 
Therefore, an overestimation of the elliptic flow cannot explain 
the observed suppression of back-to-back correlations.

Analyses of fixed-target experiments \cite{Apanasevich} have suggested
that the shape of the back-to-back dihadron azimuthal distribution is
sensitive to the intrinsic parton transverse momentum $k_T$ in the
initial state.  In proton-nucleus and nucleus-nucleus collisions,
additional initial-state transverse momentum can be generated by
multiple nucleon-nucleon interactions preceding a hard scattering
\cite{Cronin, Corcoran, Lev}.  To investigate whether this nuclear $k_T$
can account for the observed deficit of back-to-back azimuthal
correlations in central Au+Au collisions, we have carried out Pythia
\cite{lund} simulations varying the $k_T$ parameter.  A rather extreme
change from the nominal value of $\sigma = 1$ GeV/c to 4 GeV/c
introduced only a small effect, reducing the predicted
$I_{AA}(2.24,\pi)$ by less than 20\%.
Experimental study of initial state effects on the azimuthal correlations 
requires the measurement of p(d)+Au collisions at RHIC energies.

In addition to the present data, two other striking effects have been
observed at high $p_T$ in nuclear collisions at RHIC: strong suppression of the
inclusive hadron yield in central 
collisions \cite{PHENIX_highpt,STAR_highpt}, and large elliptic
flow which saturates at $p_T>3$ GeV/c \cite{STARv2}. 
These phenomena are all consistent
with a picture in which observed hadrons at $p_T>3-4$ GeV/c are fragments of
hard scattered partons, and partons or their fragments are strongly
scattered or absorbed in the nuclear medium. The observed hadrons therefore
result preferentially from hard-scattered partons generated on the periphery of the
reaction zone and heading outwards \cite{Bjorken}. 
In this picture
the inclusive yield will be suppressed relative to the binary scaling
expectation, and the strong position-momentum correlation required to
explain the large elliptic flow \cite{Shuryak} emerges naturally. 
The properties of
small-angle hadron correlations will have weak dependence on the size of the
colliding system, whereas the back-to-back correlations will exhibit strong
suppression for a large system relative to a small one, both as observed.

In summary, STAR has measured azimuthal correlations for high $p_T$
charged particles over a large relative pseudo-rapidity range with full
azimuthal angle coverage.  Comparison of the opposite-sign and same-sign
correlation strengths indicates that hard scattering and fragmentation is the
predominant source of charged hadrons with $p_T>4$ GeV/c in central Au +
Au collisions. The azimuthal correlations in Au+Au collisions have been
treated as the superposition of independently determined elliptic flow and 
individual hard parton scattering contributions, the latter measured
in the STAR p+p data.  The most striking feature of the hard-scattering
component is an increasing suppression of back-to-back relative to small-angle
correlations with increasing centrality.  These observations appear consistent
with large energy loss in a system that is opaque to the propagation
of high-momentum partons or their fragmentation products.

We wish to thank the RHIC Operations Group and the RHIC Computing Facility
at Brookhaven National Laboratory, and the National Energy Research 
Scientific Computing Center at Lawrence Berkeley National Laboratory
for their support. This work was supported by the Division of Nuclear 
Physics and the Division of High Energy Physics of the Office of Science of 
the U.S. Department of Energy, the United States National Science Foundation,
the Bundesministerium fuer Bildung und Forschung of Germany,
the Institut National de la Physique Nucleaire et de la Physique 
des Particules of France, the United Kingdom Engineering and Physical 
Sciences Research Council, Fundacao de Amparo a Pesquisa do Estado de Sao 
Paulo, Brazil, the Russian Ministry of Science and Technology and the
Ministry of Education of China and the National Natural Science Foundation 
of China.

\bibliographystyle{unsrt}

\end{document}